# Metabolic scaling is governed by Murray's network in animals and by hydraulic conductance and photosynthesis in plants

Jinkui Zhao

Songshan Lake Materials Laboratory and Institute of Physics, Chinese Academy of Sciences.

Email: jkzhao@iphy.ac.cn

**The prevailing theory for metabolic scaling is based on area-preserved, space-filling fractal vascular networks. However, it's known both theoretically and experimentally that animals' vascular systems obey Murray's cubic branching law. Area-preserved branching conflicts with energy minimization and hence the least-work principle. Additionally, while Kleiber's law is the dominant rule for both animals and plants, small animals are observed to follow the 2/3-power law, large animals have larger than 3/4 scaling exponents, and small plants have near-linear scaling behaviors. No known theory explains all the observations. Here, I show that animals' metabolism is determined by their Murray's vascular systems. For plants, the scaling is determined by the trunks' hydraulic conductance and the leaves' photosynthesis. Both analyses agree with data of various body sizes. Animals' scaling has a concave curvature while plants have a convex one. The empirical power laws are approximations within selected mass ranges. Generally, the 3/4-power law applies to animals of ~15 g to 10,000 kg and the 2/3-power law to those of ~1 g to 10 kg. For plants, the scaling exponent is 1 for small plants and decreases to 3/4 for those greater than ~10 kg.**

Kleiber's law[1] that metabolism scales with the 3/4-power of body mass has wide support in animals and plants[2,3]. Exceptions however are found at both ends of the size spectrum: small animals scale with the 2/3-power law[4–6]; large animals such as cetaceans have larger than 3/4 power exponents[7]; and small plants have a close to linear scaling[8,9]. The 2/3-power law would agree with the early surface heat dissipation hypothesis, but this is not supported by current analyses[10–22]. Currently, the most influential theory explaining Kleiber's law is based on area-preserving fractal vascular networks[23,24], though it is not universally accepted[25–28]. The issue is that it has been shown both theoretically[29,30] and experimentally[31–36] that the branching of animals' vascular systems obeys Murray's law[37]. Area-preserved network branching pattern is not energy optimal for animal vascular flows. Adding to the lack of theoretical agreement, as mentioned above, not all data agree with Kleiber's law[4–9]. A curvature exists in the scaling data[38]. All reported scaling observations[4,5,39] are for selected mass ranges and will have different scaling exponents if we change these ranges.

In this work, I derive animals' metabolism as it relates to body size and show that scaling governed by Murray's vascular network agrees with reported data. I also derive plants' scaling from the hydraulic conductance of the tree trunk and the photosynthesis of the leaves and validate the model with observations. The result shows that even though animals' and plants' metabolic scaling can be approximated to power laws within limited body sizes, they take on different forms. Animals' scaling has a negative concave curvature while plants' scaling has a positive convex curvature.

**Animals' vascular system and metabolic scaling.** Animals' metabolism is limited by the consumption of blood oxygen. In animals such as mammals and birds, blood and oxygen are carried by the vascular system whose branching pattern follows Murray's cubic branching law[31–36]. Cubic branching is energy-optimized for laminar flows[29,30]. To obtain the body mass-dependent metabolism, we make three basic assumptions[40]: (A1) Animals' vascular networks are space-filling; (A2) The dimensions of the network's terminal capillary are invariant[23,41]; (A3) Metabolic rate is proportional to the consumption of oxygen carried by the blood flow. These assumptions are generally used and accepted[23]. Assumption (A1) is the consequence that resources must be delivered to every part of an animal's body. (A2) arises because resources must be delivered down to the cellular level where physiology is essentially the same. For arguments that capillaries vary in dimensions but their total volume scales with body mass[42], it's shown in M8.1 that it doesn't change the basic formalism of the current discussion.

Assumption (A3) can be further quantified. It's generally observed that blood pressure is independent of body mass[43], but the blood oxygen pressure scales with the $-1/12$ power of body mass[42,44]. This variation will be taken into account in later calculations and will be verified against metabolic data. Variations in oxygen pressure are related to the oxygen affinity of the hemoglobin protein[45] and appear to be a size-dependent evolutionary adaptation in animals[46]. Blood pressure invariance can be considered in the following: on one hand, blood pressure needs to be sufficiently high to overcome gravitation and blood viscous resistances[45]; on the other hand, even though elevated pressure would allow for faster blood circulation and benefit metabolism, animals' heart and arterial tissues have limited strength. Higher blood pressures require significantly thicker arterial walls[45]. In the current discussion though, possible slight blood pressure variations[47] will have the same effect as that of oxygen pressure.

Laminar blood flow is governed by Poiseuille's law. In M3, we derive the flow impedance of Murray's vascular systems for animals and show that the network is impedance-matched between successive branching levels. Impedance matching is characteristic of optimized transport networks including electrical circuits, transport pipes, and indeed vascular systems[31]. One immediate outcome of the impedance-matched vascular system is that it agrees with the capillary blood slowdown observations. In M5, it's shown that the blood speed at the capillaries is $N_c^{1/3}$ times slower than that at the aorta, $N_c$ is the total number of capillaries. For $N_c \sim 10^{10}$ as in humans, the slowdown factor is about 1000, agreeing with observations[23].

For comparison, the impedance for an area-preserved, branching network is also derived in M9.1, showing that it is impedance mismatched. In an example of such networks with 10 branching segments from the aorta to the capillaries and each parent vessel branches into 3 daughter vessels, the flow resistance of the capillary segment will be 1500 times more than that of the aorta. Area-preserved branching is thus energetically unfavored and unrealistic for animal vascular networks.

Animals' aorta and main arteries are optimized for pulsatile and turbulent flows. Networks optimized for turbulent flows follow a 7/3-power branching rule[30]. The transition from turbulent to laminar flow occurs at some branching rank later. From the energy-minimization point of view, however, the optimized turbulent flow on the aorta side must match the optimized laminar flow on the capillary side of the vascular network. Analyzing either section must give the same blood and oxygen flow rate and thus be equally valid. Equivalently, blood backflows through the veins are predominantly laminar, though it's more intuitive to work with the oxygen-carrying artery flows. Here, we focus on the laminar flows at the arterioles and capillaries. The elastic aorta and the main arteries are viewed as pressure buffers for the pulsating heart. They take in blood during the rather short period of cardiac contraction, then exert constant pressure onto the arterioles and capillaries. This is reasonable also because blood pressures are typically measured at the main arteries rather than at the heart itself. Assuming the transition from turbulent to laminar flows occurs at the branching rank $K_{TL}$ of the vascular network (aorta is at rank 0), the metabolic rate $B$ as a function of body mass $M$ is derived in M4 as

$$B = \alpha M^x \log^{-1}(\beta M), \text{ with } x = {}^{11}/{}_{12} \qquad (1)$$

$\alpha$ is a proportionality constant, $\beta = m_c^{-1} n^{-K_{TL}}$ where $m_c$ is the single capillary service volume, $n$ is the branching ratio, namely a parent vessel branches into $n$ daughter ones. The factor $x = {}^{11}/{}_{12}$ comes from the above oxygen pressure discussion. Since turbulent flow plays a more prominent role for larger animals, it has been argued[23] that $K_{TL}$ should be proportional to $\log(M)$, $K_{TL} \sim \log(M)$. In M8.2, however, it's shown that such a modification does not change the above formalism except for a redefined $\beta$ constant.

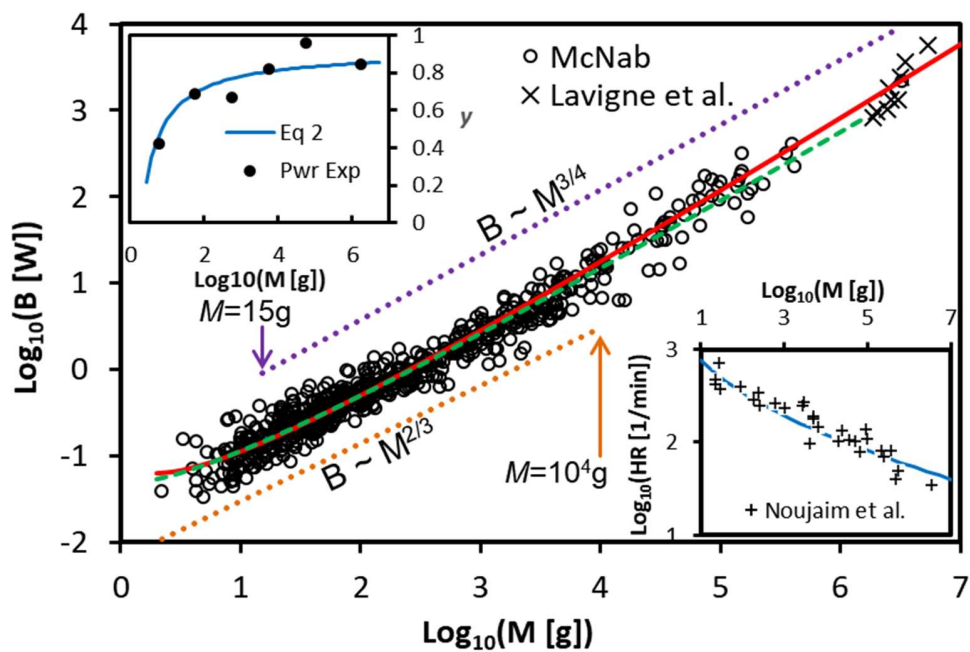

Figure 1. Equation 1 applied to BMR data[39] (circles, data range $M = 2.2 – 3 \times 10^6$ g): using $M \leq 1000$ g data only (solid red line, $\alpha = 0.01621 \pm 0.00006$ and $\beta = 1.514 \pm 0.022$); whole data set with variable $x$ (green dashed line, $x = 0.868 \pm 0.013$, $\alpha = 0.0227 \pm 0.0003$, and $\beta = 2.94 \pm 0.03$). BMR from a single whale[48] is plotted for comparison (crosses). The lines for the 3/4-power law (purple dots, valid lower mass boundary marked by purple arrow) and the 2/3-power law (orange dots, valid upper mass boundary marked by orange arrow) are shown but shifted vertically to facilitate display. Upper insert: power exponent from equation 3 (with $\beta = 1.514$ ) compared with power-law ($B \sim M^y$) fits to subsets of BMR data with the ranges of $M = 2.2$ to $10$, $10$ to $10^2$, $10^2$ to $10^3$, $10^3$ to $10^4$, $10^4$ to $10^5$, and $10^5$ to $3.22 \times 10^6$ g, plotted against each range's mass average. Lower insert: application of equation 3 (with $\beta = 1.514$) to heart rate data[49].

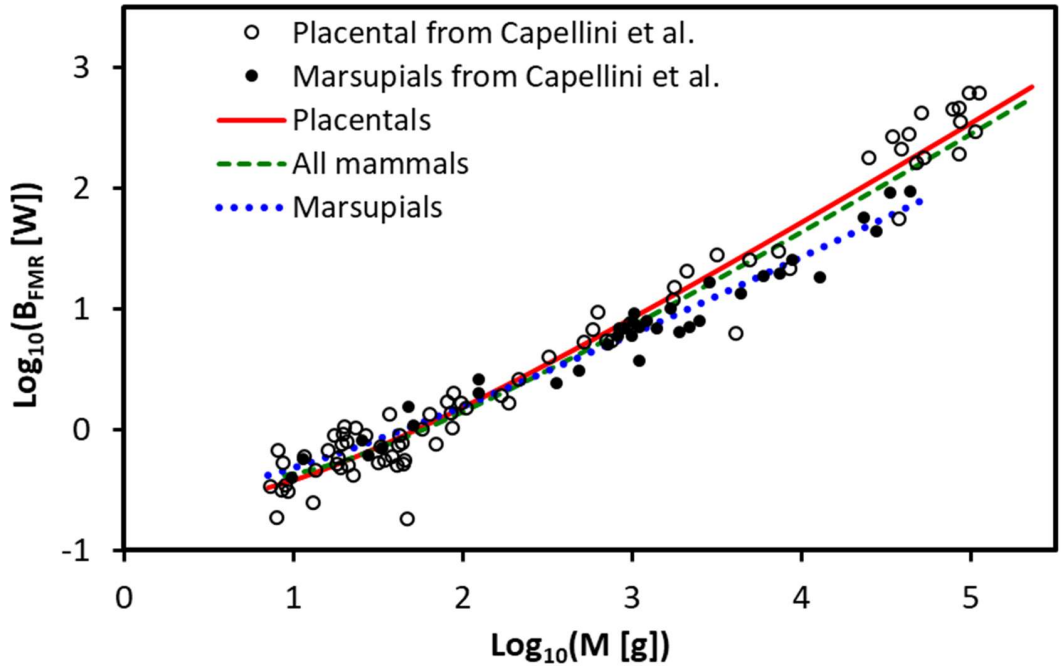

Figure 2. FMR data[5] (filled and open circles). Application of equation 1 to placental mammals (solid red line, $\alpha$ = 0.0437 ± 0.0003, $\beta$ = 0.8758 ± 0.0009), all mammals (dashed green line, $\alpha$ = 0.0336 ± 0.0002, $\beta$ = 0.451 ± 0.002), and marsupials with variable $x$ for equation 1 (dotted blue line, $x$ = 0.75 ± 0.06, $\alpha$ = 0.115 ± 0.003, $\beta$ = 2.23 ± 0.07) are shown. Allowing $x$ to vary for all mammals results in $x$ = 0.932 ± 0.026 (curve not shown), essentially the same as the $11/12$ value in equation 1.

Note that equation 1 limits $M$ to $M > \beta^{-1}$. To understand it, we look at $K_{TL} = 0$ and thus $\beta^{-1} = m_c$. $M > \beta^{-1}$ means that there should be more than one capillary in the network. For $K_{TL} \neq 0$, it means that $M$ should be large enough to allow for the pulsatile to laminar flow transition to happen. These lower mass limits on $M$ show that the scaling mechanisms for organisms without a vascular system, such as single cellular ones, must be different from that of animals with vascular systems.

The constant $\beta$ is related to the structure of the vascular system. In practice, it's convenient to determine its value from available metabolic data using equation 1. To validate this approach, we first apply equation 1 to basal metabolic rate (BMR) data[38,39] somewhat arbitrarily to $M \leq 1000$ g, then using the resulting $\beta$ = 1.514 to predict BMR for larger animals. Figure 1 shows that the prediction is very accurate within the range of available data of up to 10 tons including that of a whale[48]. It also confirms that large whales indeed have higher BMRs[7]. To make sense of the obtained value of $\beta$ = 1.514, we look at an example network with $n = 3$ and $K_{TL} = 7$. It gives a capillary length of ~ 0.7 mm, a value that matches observations[50]. $\beta$ values from different BMR datasets will reasonably differ, but they do not affect the core discussion. In the following, we will use $\beta$ = 1.514 for further discussions.

Since the $M^{-1/12}$ oxygen pressure dependence is empirical, we test it by allowing $x$ in equation 1 to vary. The test is carried out using both the BMR[39] (Fig. 1) and the field metabolic rate (FMR) data[5] (Fig. 2). Both tests return an $x$ value close to $11/12$, validating the blood pressure dependency. Incidentally, the reported FMR data[5] also includes marsupials. Test of equation 1 against marsupial FMR gives $x = 0.75$. This is interesting because it has been known that marsupials' blood oxygen pressure dependence on body mass is ~3× steeper than placentals[51], meaning that their blood oxygen pressure should scale with $M^{-3/12}$. Thus, $x$ should equal be $3/4$ (=$1-3/12$) which is our exactly result. This establishes a connection between marsupials' lower FMR allometric slope and their steeper blood oxygen pressure dependence on body mass.

The logarithmic normalization in equation 1 is an intrinsic property of linearly branching networks and cannot be simply written as a power law. However, the $\log(B)$ vs. $\log(M)$ slope averaged over a mass range can be regarded as the power-law exponent for that range. The slope is derived in M6 as

$$y = {}^{11}/_{12} - \ln^{-1}(\beta M) \tag{2}$$

Ln is the natural logarithm, $\beta = 1.514$ is used per earlier discussion. The $y$-value increases with $M$ and thus the metabolism of equation 1 has a negative concave curvature[38]. We now see why empirical scaling laws are different for different body sizes. For example, the same data from the reported 2/3-power law[4] give a power exponent of ~ 0.81 when we only consider data with $M \geq 1000$ g. BMR data we've examined [4,5,39] all have the same increased scaling exponent behavior for larger animals. Figure 1 (upper insert) shows that such mass dependency is accurately described by equation 2.

Equation 2 gives a slope value of 2/3 for M ~ 36 g and 3/4 for M ~ 266 g using the value of $\beta = 1.514$. Over the reported BMR data range[39], the average $y$-values for $M \leq 10^4$ g and $M \geq 15$ g also give 2/3 and 3/4, meaning that the 2/3- and 3/4-power law approximations can be applied to the corresponding mass range. For very large animals, for $M \geq 10^5$ g for example, the average $y$-values is 0.84, confirming that large animals such as whales indeed have higher BMRs[7].

Equation 1 can be extended to other allometries such as the heart rate. It has been known[52] that heart stroke volume scales with $M$. To deliver the appropriate amount of oxygen, the heart rate $HR$ then must follow

$$HR \sim M^{-1/12}\log^{-1}(\beta M) \tag{3}$$

Figure 1 (lower insert) shows that equation 3 agrees surprisingly well with heart rate measurements[49].

**Plants' metabolic scaling.** Plants' metabolic rate $B$ is proportional to water supply through the tree trunk and to photosynthesis at the leaves[53–55]. Biomechanical stability requires[56] that the length $l$ and radius $r$ of a tree trunk or branch must satisfy $l \sim r^{2/3}$. Since the hydraulic conductance is proportional to the cross-section of the tree trunk, we have the well-known scaling rule[53–59] of $B \sim \pi r^2 \sim M_0^{3/4}$, $M_0$ is the mass of the tree trunk ($M_0 \sim \pi r^2 l$).

Water transport up the tree is driven by diffusion, for which the optimized network has the area-preserving, square-branching pattern[29]. In M9.2, we prove that scaling resulting from an area-preserved network[24] for plants is equivalent to the above result of $B \sim M_0^{3/4}$. The mass of the tree branches scales linearly with that of the trunk and thus can be considered included in $M_0$. What is now missing in this scaling rule is the leaf mass $M_L$, which is proportional to the metabolic rate $B$, $B \sim M_L$. To include $M_L$, we rewrite $B \sim M_0^{3/4}$ and $B \sim M_L$ in revert as $M_0 = \varphi B^{4/3}$ and $M_L = \theta B$, $\theta$ and $\varphi$ are two proportionality constants. The mass is then $M = M_0 + M_L$ and we have

$$M = \theta B + \varphi B^{4/3} \tag{4}$$

The above treatment implies $M_L \sim M_0^{3/4}$, which has been pointed out and verified[54]. What differs here is the realization that the scaling rule obtained from the plant vascular network[24] is equivalent to scaling against the tree trunk and does not include the leaves. The leaves attached to the end of petioles and petiolules have more masses than tree branching patterns suggest (see M9.2), especially for small plants. Figure 3 shows that equation 4 interprets plants' respiration data[8] extremely well, including those from small plants, confirming that proper consideration of the leaves is necessary and also sufficient. For the tree roots, there also exists a linear relationship between the mass of the tree trunk and the tree roots[60]. Consequently, the whole-plant data[8] are equally well interpreted by equation 4 (Fig. 3).

Inversion of equation 4 is possible but complicated (see M7) and for the current purpose, unnecessary. As with animals, the "scaling exponent" $y$ from equation 4 is derived in M7 as

$$y = 3(4 - M_L/M)^{-1} \tag{5}$$

Figure 3 (insert) shows the excellent agreement of equation 5 compared with the varying power-law exponents (similar to animals). The $y$-value decreases with $M$ and the curvature for $B$ vs. $M$ is positive and convex. For $M > 10^4$ g, $y$ starts to approach 3/4 (Fig. 3), indicating the diminishing role leaves play for larger trees. For very small plants, however, the predicted $y$ value approaches 1, confirming observations[8,9]. Figure 4 further shows that equation 5 can accurately predict leaf percentages[54] based on metabolic data[61,62].

**Discussion.** In animals, both maintaining and pumping blood incur costs. The vascular system needs to be energy efficient and adhere to Murray's cubic branching law. In plants, external energy, the sunlight, drives transpiration at the leaves which in turn drives water diffusion through the lifeless xylem. Since energetically favored networks for diffusion indeed take the form of area-preservation, we only need to examine any one segment between the tree trunk and the petioles to determine the limitation on water flow. The tree trunk offers a straightforward way with its height-radius correlation imposed by the mechanical stability requirement. Incidentally, area preservation can also be derived under the biomechanical stability and space-filling considerations (see M9.2). An area-preserving network is equivalent to biomechanical and hydraulic conductance considerations vis-à-vis allometry.

Both the animal and plant models presented in this work agree with metabolic data of various body sizes. In both cases, metabolic scaling has a curvature and the empirical power laws are approximations within confined mass ranges. Generally, the 3/4-power law applies to animals of ~15 g to 10,000 kg and the 2/3-power law to those of ~1 g to 10 kg. For plants, the scaling exponent is 1 for small plants and decreases to 3/4 for those greater than ~10 kg.

For the animal model, capillary blood slowdown and derived information based on BMR data, such as capillary length, and heart rate, all agree with respective experimental observations, further affirming the model's validity. The model also establishes a connection between marsupials' steeper dependence of blood oxygen pressure on body mass and their lower FMR allometric slope. Equally, for plants, the model's leaf percentage prediction based on metabolic data agrees with independent leaf measurements.

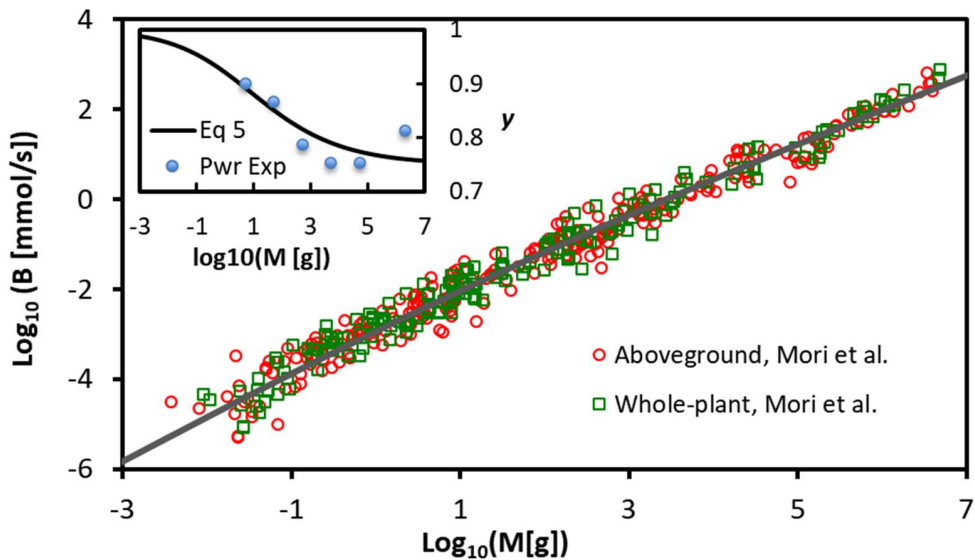

Figure 3. Equation 4 applied to fresh mass aboveground respiration data (solid line, $\theta$ = 643 ±74, $\varphi$ = 2135 ± 219). Above ground (red circles) and whole plant (green squares) data[8] are shown. Application of equation 4 to whole-plant data results in a slightly right-shifted curve (not shown) due to the added root mass. The insert shows equation 5 (see M7 for calculation) compared with power-law exponent $y$ from aboveground data at different mass ranges ($B \sim M^y$ fit to data subsets of $M = 3.8 \times 10^{-3}$ to 1, 1 to $10^2$, 10 to $10^3$, $10^2$ to $10^4$, $10^3$ to $10^5$, and $10^4$ to $4 \times 10^6$ g, respectively, plotted against the average mass of each range).

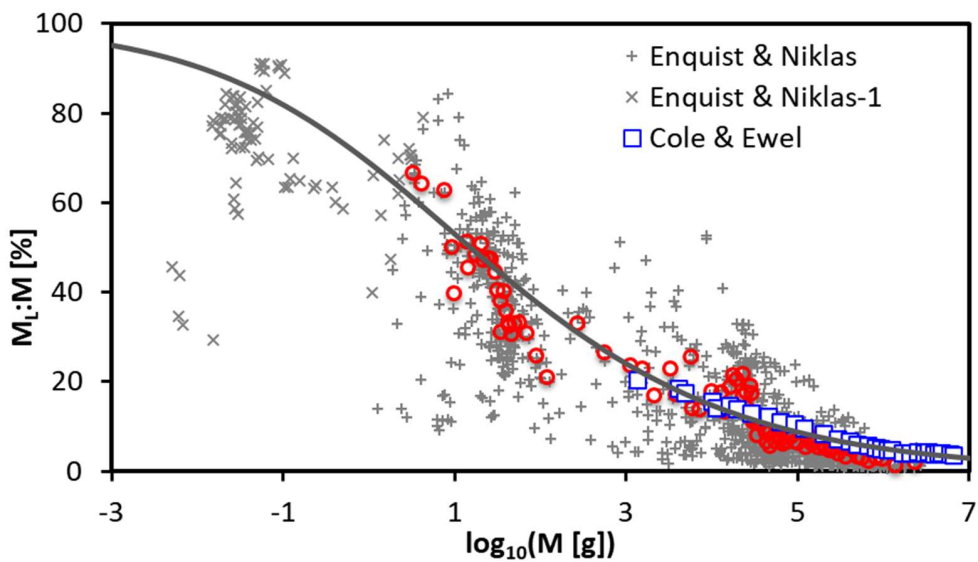

Figure 4. The $y$-curve from figure 3 is replotted as the leaf percentage (Eq. 5, solid line). The mass $M$ value is divided[62] by 2.3 to account for the difference between fresh tree masses in figure 3 and dry masses here. Leaf measurements [54] are shown as pluses and crosses. The large data dispersion is possibly due to variations among individual trees and the fact that the data contain both angiosperms and gymnosperms. A 10-point running average from the pluses to reduce the scatter follows the prediction well (red circles). The blue squares are adapted Cedrela data[61].

**Materials and Methods**

*M1. Animal vascular model.* Animals' allometry and vascular system have been extensively studied. It's been known that energy-optimized linearly branching networks under laminar flow follow Murray's branching law [29,37], which we assume here for our model. Each branching segment is denoted by a branch rank index $k$, $k = 0$ is at the aorta, and $k = K_c$ is at the terminal capillaries. The subscript $c$ denotes the capillary. $K_c$ thus also denotes the total number of branching segments from the aorta out. For the symmetric branching network where one parent vessel at rank $k$ branches into $n$ daughter vessels at rank $k + 1$, Murray's law is written as

$$r_k^3 = n r_{k+1}^3 \text{ and } r_k/r_{k+1} = n^{1/3}. \tag{6}$$

$r_k$ and $r_{k+1}$ are the vessel radii at rank $k$ and $k + 1$, respectively. At rank $k$, there are a total of $N_k = n^k$ blood vessels. At the terminal branch, the total number of capillaries is given by $N_c = n^{K_c}$. From equation 6, the radius $r_c$ and area of cross-section $s_c$ of a single capillary vessel are related to those of the aorta $r_0$ and $s_0$ by

$$r_c/r_0 = n^{-K_c/3} = N_c^{-1/3} \text{ and } s_c/s_0 = n^{-2K_c/3} = N_c^{-2/3}. \tag{7}$$

The total area of the cross-section for the capillary branching segment ($S_c = N_c s_c$) thus relates to that of the aorta ($S_0 = s_0$) by

$$S_c/S_0 = N_c^{1/3} \tag{8}$$

The length of a blood vessel $l_k$ is generally much larger than its radius, $l_k \gg r_k$. The body volume $v_k$ served by a single blood vessel should be proportional to $l_k^3$ [23] and can be simplified as $v_k = l_k^3$ for simplicity. With a constant mass density of ~ 1 g/cm$^3$ for animals, the service mass $m_k$ by a blood vessel can also be simplified as

$$m_k = l_k^3. \tag{9}$$

In the main text, the capillary service volume $m_c$ is estimated from metabolic data and the capillary length is estimated using this relationship. From the space-filling assumption (A1), a vessel and its $n$ daughter branches serve the same volume[23], namely $v_k = n v_{k+1}$, which leads to the relationship of

$$l_k/l_{k+1} = n^{1/3}. \tag{10}$$

Also from (A1), the total service mass of all capillaries $N_c m_c$ must equal to the animal's body mass $M$, namely

$$M = N_c m_c = n^{K_c} m_c. \tag{11}$$

From this, we get the total number of branching segments for a given mass $M$,

$$K_c = \log(M m_c^{-1}) \log^{-1}(n). \tag{12}$$

The capillary's service mass $m_c$ is invariant as per assumption (A2).

*M2. Animal model blood flow resistance.* Blood flows through successive network segments in serial, but within a segment, it flows through all vessels in parallel. The laminar blood flow resistance $z$ of a single blood vessel of length $l$ and radius $r$ is given by Poiseuille's law as $z \propto lr^{-4}$. With the relationships of $r_k/r_{k+1} = n^{1/3}$ and $l_k/l_{k+1} = n^{1/3}$ from the model, the flow resistances of a single blood vessel at rank $k$ and $k+1$ thus satisfy $z_k = z_{k+1}/n$. Since blood flows through all the $N_k = n^k$ vessels at rank $k$ in parallel, the segment's overall flow resistance is given by

$$Z_k = z_k/n^k. \tag{13}$$

Substitute in $z_k = z_{k+1}/n$ and we get

$$Z_k = z_{k+1}/n^{k+1} = Z_{k+1}. \tag{14}$$

Thus, the animal vascular network is impedance-matched from one branching segment to the next. The resistance of the capillary segments is calculated as

$$Z_c = z_c/N_c = z_c m_c M^{-1}. \tag{15}$$

$z_c$ and $m_c$ are invariants. $Z_c$ is thus inversely proportional to mass $M$. Using equations 11 and 12, the overall system flow resistance $Z$ is then given by the summation of all the segments $Z = Z_0 + Z_1 + \ldots + Z_c$, which equals

$$Z = K_c Z_c = K_c z_c m_c M^{-1}, \text{ and}$$

$$Z = a \log(M m_c^{-1}) M^{-1}. \tag{16}$$

where $a = z_c m_c \log^{-1}(n)$ is a constant. Note we have used $Z = K_c Z_c$ in place of $Z = (K_c + 1) Z_c$ for clarity. Treatment of the latter would follow the same steps as that for $K_{TL}$ in the next section.

***M3. Animal model flow resistance with turbulent to laminar transition.*** As discussed in the main text, we focus on the laminar flow in arterioles and capillaries. Assuming blood flow transitions from turbulent to laminar at the rank $K_{TL}$, the effective network resistance is now modified from equation 16 to $Z = (K_c - K_{TL}) Z_c$. Substitute in equation 12, we have

$$Z = \left[\log^{-1}(n) z_c m_c\right] \left[\log(M m_c^{-1}) - K_{TL} \log(n)\right] M^{-1}. \tag{17}$$

Note the logarithmic relationships of $K_{TL} \log(n) = \log(n^{K_{TL}})$ and $\log(M m_c^{-1}) - \log(n^{K_{TL}}) = \log(M m_c^{-1} n^{-K_{TL}})$, we have

$$Z = \left[z_c m_c \log^{-1}(n)\right] \log(M m_c^{-1} n^{-K_{TL}}) M^{-1}. \tag{18}$$

Next, we introduce two new constants $\alpha' = z_c m_c \log^{-1}(n)$ and $\beta = m_c^{-1} n^{-K_{TL}}$, we then have

$$Z = \alpha' \log(\beta M) M^{-1}. \tag{19}$$

Compare equation 19 to 16 where $K_{TL} = 0$, we see that the only difference is the definition of $\beta = m_c^{-1} n^{-K_{TL}}$, which would have been $\beta = m_c^{-1}$ for equation 16. Other alternative forms of $Z$ with varied capillary dimensions and mass-dependent turbulent to laminar flow transitions will be proven in section M8 to be equivalent to this form. Equation 19 will be used for metabolic rate calculation.

***M4. Animal metabolic rate.*** As discussed in the main text, the metabolic rate $B$ is proportional to oxygen pressure, which scales with $M^{-1/12}$. Thus, $B \propto M^{-1/12} Z^{-1}$. With flow resistance from equation 19, we arrive at equation 1,

$$B = \alpha M^x \log^{-1}(\beta M), \text{ with } x = {}^{11}/_{12} \tag{20}$$

and $\alpha \sim \log(n) z_c^{-1} m_c^{-1}$ is a constant that replaces $\alpha'$ from equation 19. The constant $x = {}^{11}/_{12}$ comes from the $-{}^{1}/_{12}$ oxygen pressure scaling exponent and is written as variable $x$ here. This is to allow for verification and fine-tuning. In some cases, such as Marsupials, $x$ deviates from ${}^{11}/_{12}$ because their blood oxygen pressure's dependence on body mass has a ~3× steeper slope than placental mammals[51].

For equation 20, if we introduce new variables $L_B = \log(B)$, $L_M = \log(M)$, $L_\alpha = \log(\alpha)$, and , $L_\beta = \log(\beta)$, it becomes

$$L_B = L_\alpha + x L_M - \log(L_\beta + L_M), \text{ with } x = {}^{11}/_{12} \tag{21}$$

*M5. Blood flow speed at the capillary.* In the vascular system, the blood mass is conserved and its flow speed $u_k$ at any branching rank $k$ is inversely proportional to that segment's total cross-section $S_k$, $u_k \sim S_k^{-1}$. From equation 8, we get the ratio of blood speeds at the aorta $u_0$ and at capillaries $u_c$ as

$$u_0/u_c = S_c/S_0 = N_c^{1/3}. \tag{22}$$

Because larger animals have a higher number of capillaries (note $N_c = M/m_c$), they will have a bigger blood slowdown ratio at the capillaries.

*M6. Animal power-law scaling exponent.* The logarithmic normalization in equation 1 (i.e. 20) is an intrinsic property of linearly branching networks, which means that it cannot be simply written as a power law. However, the curvature of $\log(B)$ vs. $\log(M)$ is rather gentle, we can view the average $\log(B)$ vs. $\log(M)$ slope over a data range as the power-law scaling exponent for those body mass data. From equation 20, we have $\log(B) = \log(\alpha) + ({}^{11}/_{12})\log(M) - \log(\log(\beta M))$. The slope is then given by the derivative $y = d(\log(B))/d(\log(M))$, which results into (*i.e.* equation 2),

$$y = {}^{11}/_{12} - \ln^{-1}(\beta M). \tag{23}$$

Ln is the natural logarithm. Note the base for all other logarithmic functions used in the current work can be any valid number but it's convenient to assume it to be 10.

*M7. Plant power-law scaling exponent.* Plants' metabolic scaling does not follow a simple power law either. To obtain the more commonly used form of $B$ as a function of $M$, equation 4 can be inverted by applying variable substitution. Replace $B$ with $B = Y^3$, equation 4 then becomes a quartic equation and can be solved using standard procedures. However, the result is lengthy and is not needed for the current discussion.

As with animals, the $\log(B)$ vs. $\log(M)$ slope is approximated as the scaling exponent. Apply derivative $y = d(\log(B))/d(\log(M)) = (M/B)/(dM/dB)$ to equation 4, we obtain $y = M/[\theta B + (4/3)\varphi B^{4/3}]$. Replace $B$ using the definitions of $M_L = \theta B$ and $M_0 = \varphi B^{4/3}$, we obtain equation 5,

$$y = 3(4 - M_L/M)^{-1}. \tag{24}$$

$y$ can also be expressed in the mass ratio of the trunk as

$$y = 3(3 + M_0/M)^{-1}. \tag{25}$$

To facilitate calculations, $y$, $M_L/M$, and $M_0/M$ can also be expressed in $B$ as

$$y = (\theta + \varphi B^{1/3})(\theta + (^4/_3)\varphi B^{1/3})^{-1} \text{ and} \tag{26}$$

$$M_L/M = (1 + (^\varphi/_\theta)B^{1/3})^{-1}. \tag{27}$$

$$M_0/M = (1 + (^\theta/_\varphi)B^{-1/3})^{-1}. \tag{28}$$

To plot $y$, $M_L/M$, or $M_0/M$ against $M$ as in figure 3, we first obtain the $\theta$ and $\varphi$ parameters by applying equation 4 to experimental data, then use $B$ as the variable to calculate $y$, $M_L/M$, and $M_0/M$.

***M8. Animal model modifications*** There are two alternatives to the details of our animal model. The first one is a modification to assumption (A2) that capillary dimensions $r_c$ and $l_c$ vary but the total capillary volume scales linearly with animal body mass $M$, namely $N_c r_c^2 l_c \sim M$[42]. The second one is that the turbulent to laminar blood flow transition location $K_{TL}$ depends on body mass with $K_{TL} \sim \log(M)$ [23]. We consider both of them here and prove that they do not change our discussion. Neither of them changes the basic form of flow resistance $Z$ and thus is equivalent in formalism to the above treatments.

<u>*M8.1. Varied capillary dimensions.*</u> For varied capillary dimensions [42] with $N_c r_c^2 l_c \sim M$, we note that assumption (A1) is still valid. Namely, the total body mass served by the capillaries equals to that of the animal, $N_c l_c^3 \sim M$. Combining the two proportionality relationships, we obtain $l_c \sim r_c$. With the single capillary flow resistance $z_c \sim l_c r_c^{-4}$ and the single capillary service mass $m_c = l_c^3$, the term $z_c m_c$ in equation 18 thus cancels to a constant since $z_c m_c \sim l_c^4 r_c^{-4}$. Since the capillary length $l_c$ now scales with some power of body mass [43], so should $m_c$. Therefore, we write $m_c = gM^h$ where $g, h$ are two constants. Let $q = z_c m_c$ and note the relationships of $Mm_c^{-1} = M^{1-h}g^{-1}$, $\log(M^{1-h}g^{-1}n^{-K_{TL}}) = (1-h)\log(M(gn^{K_{TL}})^{1/(h-1)})$, equation 18 now becomes

$$Z = q(1-h)\log^{-1}(n)\log(M(gn^{K_{TL}})^{1/(h-1)})M^{-1} \tag{29}$$

Next, we introduce two new constants $a' = q(1-h)\log^{-1}(n)$ and $b' = (gn^{K_{TL}})^{1/(h-1)}$, we have

$$Z = a'\log(b'M)M^{-1}. \tag{30}$$

This is the same form as equation 19. Therefore, the varied capillary dimension case does not change the basic form of flow resistance $Z$ and hence will not affect the equation for metabolic rate $B$.

<u>*M8.2. Mass-dependent turbulent to laminar transition.*</u> For the second case where $K_{TL}$ varied[23] with $\log(M)$, we replace $K_{TL}$ in equation 17 with

$$K_{TL} = k'\log(M), \tag{31}$$

$k'$ is a proportionality constant. Again, substitute equation 12 into equation 17, we obtain

$$Z = z_c m_c M^{-1}\left[\log(Mm_c^{-1})\log^{-1}(n) - k'\log(M)\right]$$

$$= z_c m_c M^{-1}\left[(\log^{-1}(n) - k')\log(M) + \log(m_c^{-1})\log^{-1}(n)\right]$$

$$= z_c m_c M^{-1}(\log^{-1}(n) - k')\left[\log(M) + \log(m_c^{-1})(1 - k'\log(n))^{-1}\right]$$

$$= z_c m_c M^{-1}(\log^{-1}(n) - k')\left[\log(M) + \log(m_c^{1/(k'\log(n) - 1)})\right] \tag{32}$$

Define constants $a''$ and $b''$ as

$$a'' = z_c m_c \left(\log^{-1}(n) - k'\right), \text{ and } b'' = m_c^{1/(k'\log(n) - 1)} \tag{33}$$

we obtain

$$Z = a''\log(b''M)M^{-1}. \tag{34}$$

Thus, other than the redefined new constants $a''$ and $b''$, the flow resistance $Z$ remains the same as equation 19 from the constant $K_{TL}$ discussion. The metabolic rate $B$ of equation 1, therefore, remains unchanged.

Following the same steps as above, it can be shown that the combination of varied capillary dimensions and varied turbulent to laminar transition can be similarly treated and the result will not change the current discussion.

***M9. Alternative, area-preserving networks*** For the alternative, often-cited area-preserved network branching model[23], we show here that it is not impendence matched for laminar flows when applied to animals and thus should not be valid. When applied to plants[24], it is equivalent to biomechanical and hydraulic conductance considerations and thus is incomplete.

<u>*M9.1. Impedance mismatch when applied to animals.*</u> For the area-preserved branching model[23], the branching structure is $r_k^2 = n r_{k+1}^2$, namely the total areas of the daughter vessels equal that of the parent vessel. Here the space-filling assumption is still valid[23], namely $l_k/l_{k+1} = n^{1/3}$. Applying Poiseuille's law, we obtain the relationship of single vessel flow resistance $z$ at rank $k$ and $k+1$,

$$z_k = z_{k+1}/n^{5/3}. \tag{35}$$

The overall flow resistance of all vessels in the rank $k$ segment is given by $Z_k = z_k/n^k$. Thus, we have

$$Z_{k+1} = n^{2/3} Z_k. \tag{36}$$

This means that the impedance is not matched from one segment to the next. Between the aorta and the capillaries, we have $Z_c = n^{2K_c/3} Z_0$. For an example network of $n = 3$ and $K_c = 10$, the flow resistance of the capillary segment is 1516 times that of the aorta, $Z_c \approx 1516 Z_0$.

<u>*M9.2. Incomplete when applied for plants.*</u> Now we look at the case when the area-preserving network is applied to plants[24]. Again, using the area-preservation and spacing-filing rules, $r_k^2 = n r_{k+1}^2$ and $l_k/l_{k+1} = n^{1/3}$, we get

$$v_k/v_{k+1} = n^{4/3} \tag{37}$$

where $v_k = \pi r_k^2 l_k$ is the volume of a single branch at rank $k$. The total volume of all branches at rank $k$ is then $V_k = n^k v_k$ and is related to that of the tree trunk $V_0$ at $k = 0$ by

$$V_k = n^{-k/3} V_0. \tag{38}$$

The same equation can be written for masses $M_k$ since tree mass density is a constant,

$$M_k = n^{-k/3} M_0. \tag{39}$$

The total tree mass $M$ is then given by

$$M = \Sigma M_k = M_0/(1 - n^{-1/3}), \tag{40}$$

namely the tree mass $M$ scales linearly with that of the trunk $M_0$. Thus, the $B \sim M^{3/4}$ relationship as derived in ref. [24] is equivalent to $B \sim M_0^{3/4}$. The geometric series summation $\Sigma M_k$ above is essentially the same as that of $V_w$ in In box 3 of ref. [24], which assumes that the number of branch ranks $K_c$ is very large such that $n^{(K_c+1)/3} \gg 1$. Here, for notation consistency with our animal model discussion, we use subscript $c$ to denote the petiole segment and $K_c$ as the total number of branch segments from the tree trunk to the petioles.

Note that leaves attached to the end of petioles and petiolules are not factored in the above tree branching pattern and thus are not taken into account. This means that the above metabolic scaling relationship derived in ref [24] does not include the mass of the leaves.

A more succinct way to arrive at the above conclusion is by analyzing the tree branching network structure. From equation 37, we obtain the tree trunk volume's relationship to that of a single petiole $v_c$ as

$$V_0 = n^{K_c 4/3} v_c. \tag{41}$$

$n^{K_c}$ is the number of petioles and is proportional to photosynthesis and thus proportional to the metabolic rate $B$. Therefore, we have

$$B \sim n^{K_c} \sim V_0^{3/4} \sim M_0^{3/4}. \tag{42}$$

As a last note, area-preserved branching patterns for plants can also be derived from the biomechanical stability requirement of $l \sim r^{2/3}$ (see main text) and the space-filling assumption of $l_k/l_{k+1} = n^{1/3}$. Since $r_k/r_{k+1} = (l_k/l_{k+1})^{3/2} = n^{1/2}$, we thus have $r_k^2 = n r_{k+1}^2$, which is the area-preservation equation with branching number $n$. In fact, any one of the three conditions (biomechanical stability $l_k \sim r_k^{2/3}$, area-preservation $r_k^2 = n r_{k+1}^2$, and spacing-filling $l_k/l_{k+1} = n^{1/3}$) can be derived from the other two.